\documentclass[]{aa}  
\usepackage{graphicx}
\usepackage{longtable}
\usepackage{txfonts}

\usepackage{natbib}

\begin{document}

\title{Modelling solar-like variability for the detection of Earth-like planetary transits}
\subtitle{II. Performance of the three-spot modelling, harmonic function fitting, iterative non-linear filtering and sliding boxcar filtering }

    \titlerunning{Solar-like activity and planetary transits}
    \authorrunning{A. S. Bonomo et al.}


   \author{A.~S.~Bonomo\inst{1,2,3}, S.~Aigrain\inst{4}, P.~Bord\'e\inst{5} \& A.~F.~Lanza\inst{2}}

   \offprints{A. S. Bonomo}

   \institute{Dipartimento di Fisica e Astronomia, Sezione Astrofisica, Universit\`a degli Studi di Catania, Italy \\
             \email{aldo.bonomo@oact.inaf.it} \and
             INAF-Osservatorio Astrofisico di Catania, Via S. Sofia, 78 -- 95123 Catania, Italy, \and
             Laboratoire d'Astrophysique de Marseille (UMR 6110), 38 rue F.~Joliot-Curie, 13388 Marseille cedex 13, France,  \and
             School of Physics, University of Exeter, Stocker Road, Exeter EX4 4QL, UK, \and
             Institut d'astrophysique spatiale, b\^at. 121, Universit\'e Paris-Sud 11 and CNRS (UMR 8617), 91405 Orsay, France.
             }

   \date{Received... ; accepted... }

    \abstract
{}{As an extension of a previous work, we present a comparison of four methods of filtering solar-like variability to increase 
the efficiency of detection of Earth-like planetary transits by means of box-shaped transit finder algorithms. Two of these filtering methods are the harmonic fitting method and the iterative non-linear filter that, coupled respectively with the Box Least-Square (BLS) and Box Maximum-Likelihood algorithms, demonstrated the best performance during the first detection blind test organized inside the CoRoT consortium. The third method, the 3-spot model, is a simplified physical model of Sun-like variability and the fourth is a simple sliding boxcar filter.}
{ We apply a Monte Carlo approach by simulating a large number of 150-day light curves (as for CoRoT long runs) for different planetary radii, orbital periods, epochs of the first transit and standard deviations of the photon shot noise. Stellar variability is given by the Total Solar Irradiance variations as observed close to the maximum of solar cycle 23. { After filtering solar variability, transits are searched for by means of the BLS algorithm.}} 
{We find that the iterative non-linear filter is the best method to filter light curves of solar-like stars when a suitable window can be chosen. As the performance of this filter depends critically on the length of its window, we point out that the window must be as long as possible, according to the magnetic activity level of the star. We show an automatic method to choose the extension of the filter window from the power spectrum of the light curves.}
{The iterative non-linear filter, when used with a suitable choice of its window, has a better performance than more complicated and computationally intensive methods of fitting solar-like variability, like the 200-harmonic fitting or the 3-spot model.}

\maketitle
%

\section{Introduction}
To date more than 50 transiting planets over about 300 known extrasolar
planets have been discovered\footnote{See http://exoplanet.eu/}. They are the most
interesting to study since transits remove the degeneracy \-bet\-we\-en orbital 
inclination and amplitude of the radial velocity curve providing us with information
on planetary masses and radii. Only for transiting planets it is possible 
to develop accurate models of their internal structures, study their 
atmospheres through transmission spectroscopy or infrared emission and
detect possible spin-orbit misalignments by measuring the Rossiter-McLaughlin effect.

The new frontier is the search for transiting terrestrial pla\-nets. Very recently 
four super-Earths have been discovered, three of them by means of the radial velocity 
technique \citep{Mayoretal08} and one by microlensing \citep{Bennettetal08}. The latter 
is a $3.3$ Earth-mass planet and is \-pro\-ba\-bly the lowest mass exoplanet found to date, 
apart from planets around pulsars \citep{WolszczanFrail92, Wolszczan94}. To detect 
the transits of terrestrial \-pla\-nets we need to go to space since the photometric 
precision from the ground is limited to the millimagnitude level because of the 
scintillation effects produced by the Earth's atmosphere. The CoRoT space mission, 
currently operating, can reach the \-pho\-to\-me\-tric precision to detect transits of 
Earth-size planets in short period orbits around solar-like stars \citep{Baglin03, Bordeetal03}.
The Kepler space mission, whose launch is planned in spring 2009, has been 
specifically designed to find also Earthly twins, i.e., Earth-size planets in 
Earth-like orbits, transiting solar-type stars \citep{Boruckietal04}.

One common approach to searching for planetary transits in light curves of main-sequence 
late-type stars  requires two main steps: first, the filtering of stellar va\-ria\-bi\-li\-ty 
to remove distortions produced by the presence of photospheric cool spots and bright 
faculae, whose visibility is modulated by stellar rotation; secondly, the search for 
transits in the filtered light curve by means of suitable detection algorithms. The 
intrinsic stellar variability represents the main source of astrophysical noise in the 
detection of transits of terrestrial planets even in relatively inactive stars 
\citep[cf., e.g., ][]{Defayetal01, Jenkins02, AigrainIrwin04, Aigrainetal04}.

Several methods to reduce the impact of solar-like \-va\-ria\-bi\-li\-ty on 
the detection of planetary transits have been developed 
\citep [see e.g.] [] {Defayetal01, Jenkinsetal02, Carpanoetal03, AigrainIrwin04, 
Moutouetal05, Reguloetal07}. Some of them were tested on simulated light curves 
during the first CoRoT blind test carried out by \citet{Moutouetal05}. The best 
performance in terms of reduction of missed detections and false alarms was 
achieved by team 3 who made use of a linear combination of 200 harmonic 
functions to filter stellar variability and the Box-fitting Least Square (BLS) 
algorithm by \citet{Kovacsetal02} to search for transits in the filtered light curves. 
Team 5 got the second best performance using the iterative non-linear filter 
by \citet{AigrainIrwin04} in combination with the Box Maximum Likelihood transit 
finder algorithm (see \citealt{AigrainIrwin04}, Sect. 2).

In a recent paper, \citet{BonomoLanza08} proposed a different method to treat 
stellar variability, called the 3-spot model, based on the rotational modulation 
of the flux produced by three point-like active regions \citep{Lanzaetal03, Lanzaetal07}. 
By analysing a large number of simulated light curves with photon noise, 
Sun-like variability and planetary transits, they compared its performance 
with that of the 200-harmonic fitting \-u\-sing the same transit detection 
algorithm (BLS), to search for transits after the filtering process. They 
found that the 3-spot model has a better performance than the 200-harmonic 
fitting when the standard deviation of the noise is 2--4 times larger than 
the central depth of the transits. On the other hand, the 200-harmonic fitting 
reduces more efficiently the impact of stellar variability when the standard 
deviation of the noise is comparable to the transit depth. 
\citet{BonomoLanza08} showed that the poor performance of the 200-harmonic 
fitting in the former case is due to the use of orthogonal functions to fit 
stellar variability, which makes it significantly affected by the Gibbs 
phenomenon \citep{MorseFeshbach54}. This latter reduces the depth of the transits 
in the filtered light curves thus lowering the efficiency of detection in the 
presence of noise (see \citealt{BonomoLanza08}, Sect.5).

The Gibbs phenomenon also affects other filtering methods proposed to reduce the impact
of stellar microvariability, such as, e.g., those of \citet{Carpanoetal03}, or the Wiener-like
discrete filters by \citet{AigrainIrwin04}. Wavelet-based methods 
\citep[e.g., ][]{Jenkins02, Reguloetal07} may be useful to overcome the problems 
related to the Gibbs phenomenon, given the non-orthogonal nature of their basis 
functions, but they are negatively affected by gaps in the time series. 
On the other hand, the iterative non-linear filter proposed by \citet{AigrainIrwin04} 
is practically insensitive to gaps or irregular sampling and does not make 
use of orthogonal functions. Therefore, in the present paper, we extend our 
comparison to it and to another simple sliding boxcar filter which has been 
recently applied by, e.g., \citet{Bordeetal03} and \citet{CarpanoFridlund08}.

\section{Filtering methods}
\label{methods}

A brief description of the filtering methods whose performance will be compared is as follows:

\begin{enumerate}

\item[a)] \emph{3-spot model} \citep{Lanzaetal03, Lanzaetal07}: it is a simplified physical model 
of solar-like variability based on the rotational modulation of the flux produced 
by three active regions, containing both cool spots and warm faculae, plus 
a uniform background to account for uniformly distributed active regions. 
In the case of the Sun, the model accounts for the flux variability up to 
a time scale of 14 days, after which the positions and areas of the three 
regions and the uniform background have to be changed (for further details 
see \citealt{BonomoLanza08}, Sect.3.1);

\item[b)] \emph{200-harmonic fitting} (\citealt{Moutouetal05}, team 3): it fits stellar 
variability by means of a linear combination of 200 harmonic functions 
whose frequencies are multiples of the fundamental frequency $f_{\rm L}=\frac{1}{2T}$, 
where $T$ is the whole duration of the time series, i.e., $T\sim150$ days 
in the case of the CoRoT mission (for further details see \citealt{BonomoLanza08}, Sect.3.2);

\item[c)] \emph{Sliding boxcar filter} (hereinafter SBC filter): it is based on 
a continuum computed by means of a running average of the data with a boxcar 
window of fixed duration, i.e., each point of the original light curve is 
replaced by the arithmetic mean of the data points falling within a boxcar 
window centered at that point. Since at the beginning and the end of the light 
curve there are not enough data points within the window to compute the continuum, 
the light curve is extended by \-mir\-ro\-ring the initial and the last 
data (edge reflection). The \-con\-ti\-nu\-um computed according to these 
prescriptions is then subtracted from the original light curve and the 
residuals are analysed to detect transits;

\item[d)] \emph{Iterative non-linear filter} (hereinafter INL filter; \citealt{AigrainIrwin04}; 
\citealt{Moutouetal05}, team 5): it considers a 
sli\-ding boxcar window of fixed duration and computes an 
initial continuum by \-re\-pla\-cing each point of the original light 
curve with the median of the data within the boxcar window centered 
at that point (the edges of the light curve are dealt with the technique 
of edge reflection as in the case of the SBC filter). It computes the 
residuals between the original light curve and the continuum and estimates 
their standard deviation $\sigma$ from the Median Absolute Deviation (MAD) 
as $\sigma$=1.4826 MAD. In the residual time series, the points whose absolute 
deviation is $\geq 3 \sigma$ are flagged, and the continuum is recomputed from 
the original time series without the flagged points, \-i\-te\-ra\-ting the process 
up to convergence. The final continuum is then subtracted from the original light curve.

\end{enumerate}

\section{Light curve simulation and analysis}
\label{sim_anal}

We apply a Monte Carlo approach by simulating a large number 
of light curves with sampling of 1 hour and duration 150 days 
(the extension of the CoRoT long runs) for different values of 
planetary radius $R_{p}$ ranging from 1.0 to 2.0 Earth radii, orbital
period $P$ between 5 and 50 days, and standard deviation of the photon
shot noise $\sigma$ from 100 to 1000 parts per millions (ppm). A noise
level $\sigma$=100 ppm is obtained for a star of magnitude $V\sim$12 
observed in white light by CoRoT with 1 hr integration time, while 
$\sigma$=200, 300 and 1000 ppm correspond to stars of $V\sim$13, 14 and
16, respectively, observed with the same instrument and 1 hr 
integration time\footnote{\mbox{http://corotsol.obspm.fr/web-instrum/payload.param/index.html}}. 
The phase of the first transit is taken from a uniform random distribution. 
The star is assumed to have the solar radius and mass. 
We add stellar variability, assumed in all cases to be given by 
the Total Solar Irradiance variations as observed close to the maximum 
of solar cycle 23 (e.g., \citealt{FrohlichLean04}). To reduce systematic 
effects associated with a particular realization of the TSI, we select a time 
series of the TSI of duration 150 days with a random initial 
epoch between January 1$^{\rm st}$ and January 20$^{\rm th}$ 2001. 
For each set of planetary parameters and noise level, 
we simulate 100 light curves with different noise and activity
realizations, for a total of 8000 light curves. For further details, 
see \citet{BonomoLanza08}, Sect. 2.

After filtering solar variability with the four different fil\-te\-ring methods, 
transits are searched for in the filtered light curves by means of the BLS algorithm 
(\citealt{Kovacsetal02}). See \citet{BonomoLanza08}, Sect. 3.3, for the choice 
of the algorithm's free parameter ranges and steps. The period search 
is carried out between 1 and 150 days so that also single transit events are searched for.
For this reason, as a detection statistic, we prefer to use the 
signal-to-noise ratio $\alpha$ of a putative transit (see \citealt{Kovacsetal02}, Eq.~11) 
instead of the Signal Detection Efficiency (hereinafter SDE, see \citealt{Kovacsetal02}, Eq.~6)
used, e.g., by team 3 for the analysis of the first CoRoT blind test light
curves (see Sect.~\ref{alpha_vs_SDE} for a discussion about the difference between
 these statistics).

Transitless light curves obtained by combining solar irradiance variations 
with random sequences of white Gaussian noise are analysed in the same way 
to establish the transit detection threshold $\alpha_{t}$ for each filtering method. 
For our analysis, this threshold is chosen in such a way as to have a false 
alarm rate of 1\% in a set of one hundred light curves.

The large computational load of our experiment is \-ma\-na\-ged by running our 
analyses on the grid infrastructure of the project PI2S2, allowing us to use 
up to about 2000 CPUs in parallel (e.g., \citealt{Becciani07}). The CPU time for 
filtering stellar microvariability with the four methods is on the average 
about 16 minutes, while about 3 days of elapsed time have been ne\-ces\-sa\-ry 
to analyse the complete set of 8000 light curves.

\section{Results}
\label{results}
After a series of tests, we have found that an appropriate window both for the 
SBC and the INL filters is 2 days when we adopt the TSI time series as a proxy 
for the variability of solar-like stars. Specifically, we found that the shorter 
the window, the greater the reduction of the transit depth and the worse the 
detection performance by means of the BLS algorithm. On the other hand, with a 
window longer than 2 days, spurious transit-like features might appear in the residuals, 
owing to a worse \-fil\-te\-ring of the variability.

\begin{figure*}
\centerline{
\hspace*{-2.0 cm}
\begin{minipage}[t]{5.8cm}
\begin{center}
 \includegraphics[width=6.5cm,angle=90]{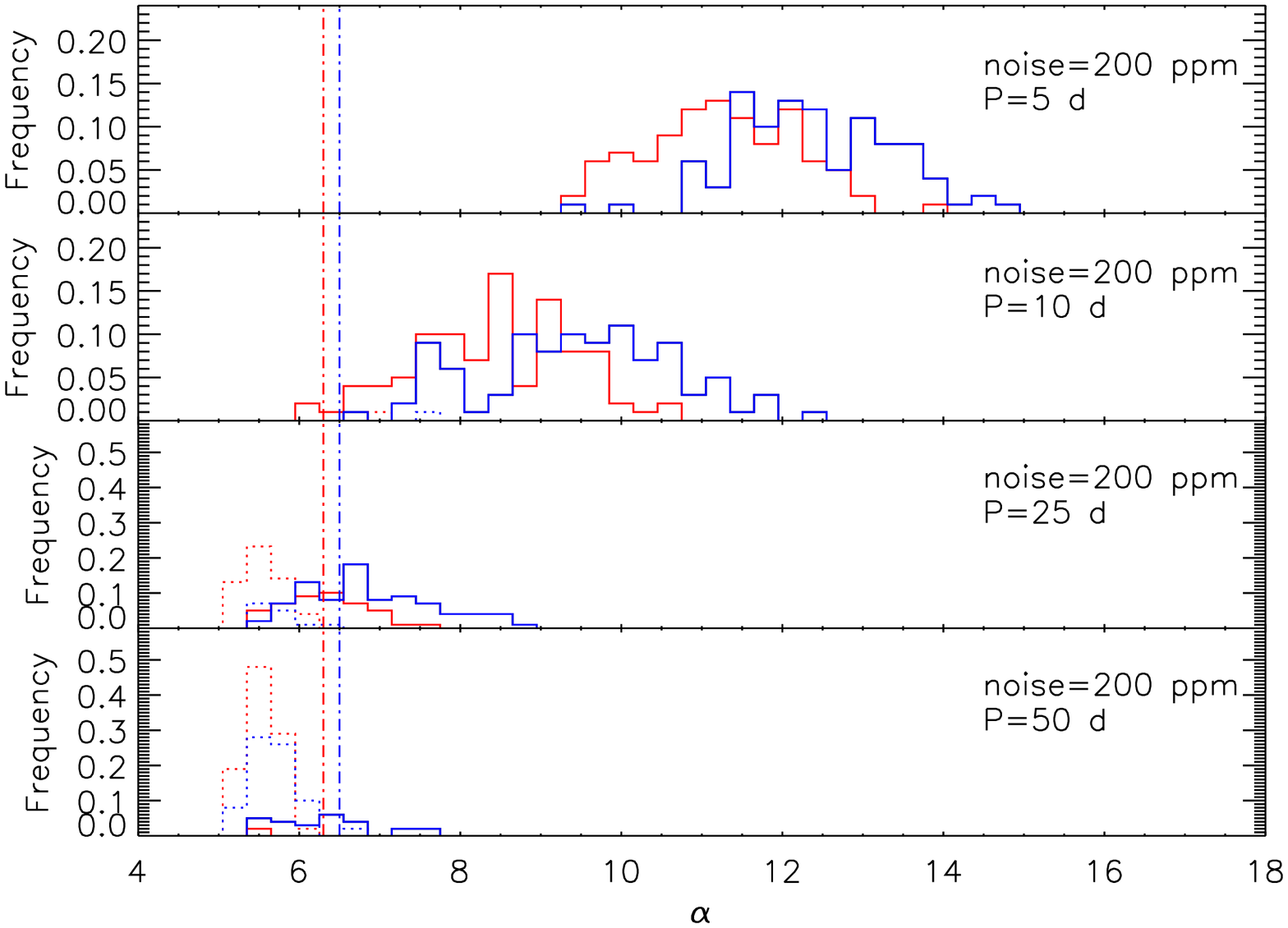} 
\end{center}
\end{minipage}
\hspace*{3.0cm}
\begin{minipage}[t]{5.8cm}
\begin{center}
 \includegraphics[width=6.5cm,angle=90]{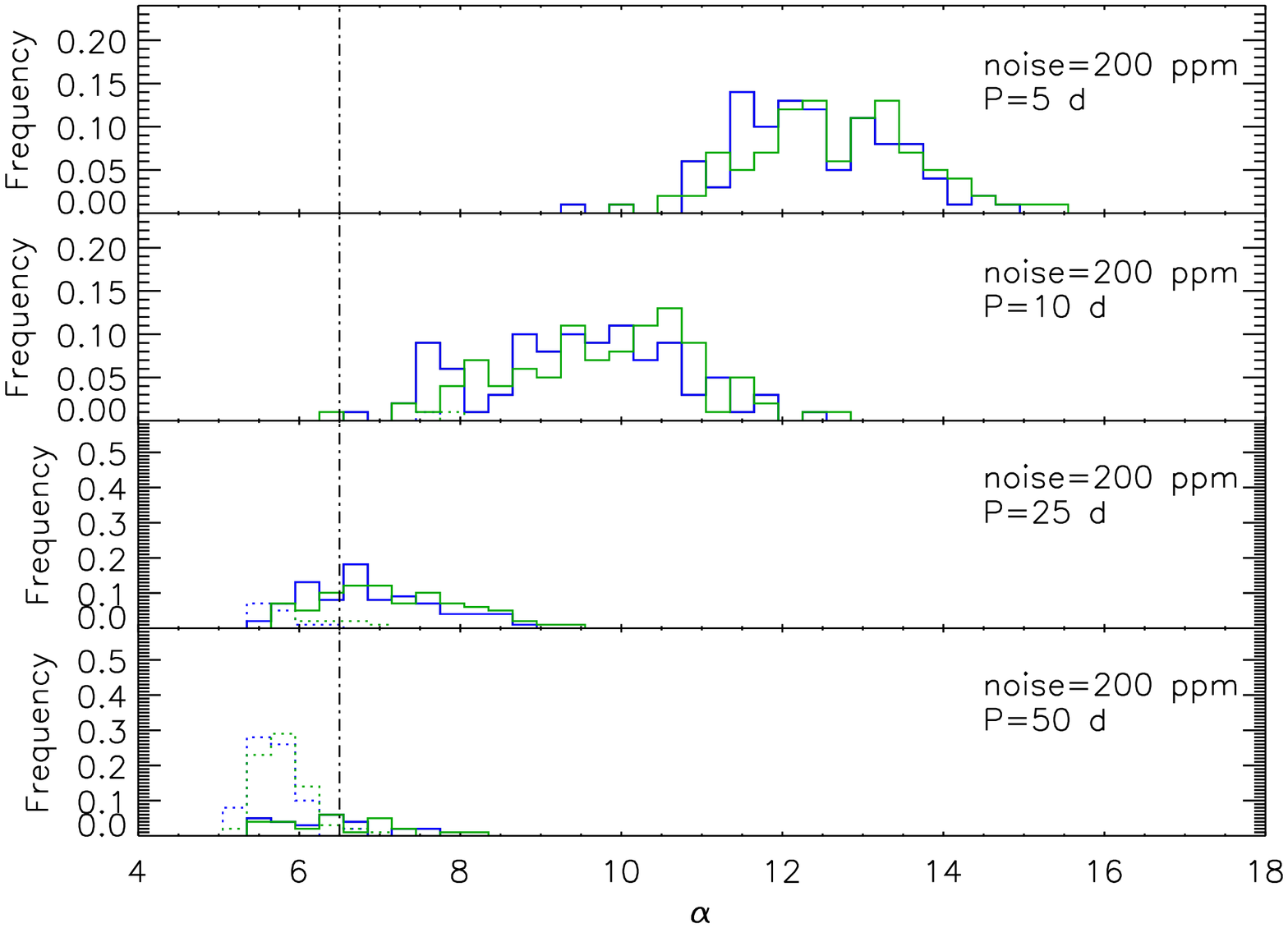}
\end{center}
\end{minipage}
}

\begin{center}
\caption{\emph{Left panels}: Distributions of the values of $\alpha$, 
the signal-to-noise ratio of a transit detection, obtained by filtering 
simulated light curves with transits of a 1.75 R$_{\oplus}$ planet, the 
labelled values of the orbital period $P$ and a standard deviation of 
the noise $\sigma=200$ ppm, by means of the 200-harmonic method 
(red solid histograms) and the 3-spot model (blue solid histograms). 
Dotted histograms refer to the cases with an incorrect period determination, 
with the same color coding. The vertical dot dashed lines indicate the 
detection thresholds, with the same color coding as the histograms 
(see the text for details). \emph{Right panels}: As on the left panels, 
distributions of the $\alpha$  values obtained analysing the simulated 
light curves after applying the 3-spot model (blue histograms) and the 
INL filter (green histograms). See text for explanation.}
   \label{fig1}
\end{center}
\end{figure*}

A sample of our results is provided in Fig.~\ref{fig1} that shows the distributions 
of the values of $\alpha$, the signal-to-noise ratio of a transit detection, obtained 
by analysing light curves with transits of a 1.75 R$_{\oplus}$ planet and white Gaussian 
noise of 200 ppm. Each set is characterized by a different orbital period $P$ of the planet 
(as labelled). 
In the left panels, the red and blue vertical dot-dashed lines indicate the 1 percent 
false-alarm threshold for the 200-harmonic fitting and the 3-spot model, respectively. 
They are derived from the analysis of the transitless light curves under the requisite 
that the frequency of false alarms be $\leq 1\%$. Red solid histograms show the statistics 
of light curves where the period $P$ was correctly identified by the BLS, within $\pm 0.1$ days, 
after applying the 200-harmonic method; blue solid histograms show those after 
applying the 3-spot method. Dashed histograms refer to the statistics of light 
curves where the period $P$ was incorrectly identified by the BLS after the 
filtering process, with the same color coding. These latter increase with 
increasing orbital period, because the number of transits in the light curve 
becomes smaller giving a lower signal-to-noise ratio. 
Right panels show the distributions of the $\alpha$ values obtained analysing the 
same simulated light curves after applying the 3-spot model (blue histograms) 
and the INL filter (green histograms). In this case, the dot-dashed vertical 
lines indicating the false alarm thresholds overlap.

Table~\ref{det_freq} reports the frequencies of detections and false alarms 
obtained for different values of planetary radius, orbital period and noise 
level, omitting the entries corresponding to null detections, with the exception 
of the first ones. As a detection, we count each case with a period correctly 
identified by the BLS and a corresponding $\alpha$ value greater than the detection 
threshold. False alarms are those cases where the $\alpha$ values are above the 
detection threshold but the period is not correctly found.
In the first column of Table~\ref{det_freq}, we list the \-pla\-ne\-ta\-ry radius $R_{p}$; 
in the se\-cond, the standard deviation of the Gaussian photon shot noise $\sigma$; 
in the third, the orbital period $P$; from the fourth to the seventh column, 
the frequency of detections obtained with the 200-harmonic fitting $D_{h}$, 
the 3-spot model $D_{3s}$, the INL filter $D_{INL}$ and the SBC filter $D_{SBC}$, 
respectively; from the eighth to the eleventh column, the frequency of false 
alarms $FA_{h}$, $FA_{3s}$, $FA_{INL}$ and $FA_{SBC}$, produced by the four filtering methods.

We report below on the performance of the different filters, according to the 
different values of the standard deviation $\sigma$ of the photon shot noise.


\subsection {$\sigma=$100 ppm}
\label{sigma_100}
\emph{Detection performance}:
The filtering methods that achieve the best performance are the INL filter and 
the 200-harmonic fitting. In most cases they give comparable results, 
although in some instances the INL filter has a slightly better performance 
(see Table~\ref{det_freq}), owing to the fact that the 200-harmonic fitting 
is affected by the Gibbs phenomenon (\citealt{MorseFeshbach54}, \citealt{BonomoLanza08}, Sect. 5).

The method that has the worst performance is the SBC filter because it gives 
rise to the greatest number of false alarms when applied to the transitless light curves. 
Therefore, we are forced to increase the transit detection threshold to push the 
false alarm rate below the 1\% level. Inspecting the light curves without transits 
that give rise to false detections, we find that they are affected by the presence 
of a positive spike close to the beginning of their 150-day TSI subsets. 
This spike is \-pro\-ba\-bly due to an instrumental effect of the VIRGO radiometers. 
When we apply the technique of edge reflection to deal with the first sections of 
these light curves (i.e., the first 24 data points), the computation of the 
continuum by means of a running \-a\-ve\-ra\-ge of the data is significantly affected 
by that spike giving rise to a transit-like feature at the beginning of the 
residual time series. The BLS algorithm detects this transit-like feature, possibly 
phased with some other residual dips, as a transit and gives rise to a false alarm. 
We could overcome such a problem by making all the points within the first day equal 
to zero, i.e., without using the edge reflection technique to deal with the beginning 
of the light curves. In this case, however, we could loose some transits whose initial 
phase is close to zero, again having a deterioration of the detection performance.

Although, in our particular case, the technique of edge reflection clearly amplifies 
the above-mentioned effect, generally speaking, a variability filter based on the 
computation of the continuum by means of a running average is certainly more 
sensitive to the presence of outliers than a median filter. This makes a median 
filter less affected by false alarms.

\vspace{0.35cm}
\noindent
\emph{False alarms}: Once the detection threshold was established by 
the analysis of the transitless light curves, the percentage of false 
alarms given by the four filtering methods in the analysis of the light 
curves with transits is usually below 2\%. Only the 200-harmonic fitting, 
in some cases, gives rise to a 4\% false alarms (see Table~\ref{det_freq}).

\subsection{$\sigma\geq$200 ppm}
\emph{Detection performance}:
The method with the best performance proves to be the INL filter when 
we use an appropriate window of 2 days. It has a performance comparable 
with that of the 3-spot model in most of the cases, even better in some 
instances (see Fig.~\ref{fig1} and Table~\ref{det_freq}). On the other hand, 
the 200-harmonic fitting has the worst performance owing to the Gibbs 
phenomenon (see Fig.~\ref{fig1} and Table~\ref{det_freq}). Finally, the SBC 
filter has a performance better than the 200-harmonic fitting but slightly 
worse than the 3-spot model and the INL filter for the same reason 
explained above, i.e., the transit detection threshold (or $\alpha$ threshold) 
must be increased with respect to those of the other filtering methods to 
maintain the false alarm rate below 1\% in the transitless light curves. 

\vspace{0.35cm}
\noindent
\emph{False alarms}: Once the detection threshold was established by the 
analysis of the transitless light curves, the percentage of false 
alarms given by the four filtering methods in the light curves with 
transits is usually below 3\%. None of the four filters produces significantly 
more false alarms than the others (see Table~\ref{det_freq}).

\begin{table*}
\caption{Fraction of positive detections and false alarms obtained in our experiment.}
\begin{center}
\begin{tabular}{|ccc|cccc|cccc|}
 \hline
& & & & & & & & & & \\
$R\mbox{~~}$ & $\sigma\mbox{~~~}$ & $P$ & $D_{\rm h}$ & $D_{\rm 3s}$ & $D_{\rm INL}$ & $D_{\rm SBC}$ & $FA_{\rm h}$ & $FA_{\rm 3s}$ & $FA_{\rm INL}$ & $FA_{\rm SBC}$ \\
  ($R_{\oplus}$) & (ppm) & (d) & & & & & & & & \\ 
& & & & & & & & & & \\
\hline
& & & & & & & & & &   \\
1.0 & 100  & 5.0   & 0.93  & 0.36  & 0.91  & 0.14  & 0.00  & 0.01  & 0.00  & 0.00 \\
    &      & 10.0  & 0.31  & 0.03  & 0.31  & 0.00  & 0.04  & 0.00  & 0.02  & 0.00 \\
    &      & 25.0  & 0.01  & 0.00  & 0.01  & 0.00  & 0.01  & 0.00  & 0.00  & 0.01 \\
    &      & 50.0  & 0.00  & 0.00  & 0.00  & 0.00  & 0.01  & 0.01  & 0.00  & 0.00 \\
1.0 & 200  & 5.0   & 0.02  & 0.06  & 0.08  & 0.06  & 0.01  & 0.01  & 0.01  & 0.01 \\
    &      & 10.0  & 0.00  & 0.00  & 0.00  & 0.00  & 0.00  & 0.00  & 0.00  & 0.01 \\
& & & & & & & & & &   \\
1.25 & 100 & 5.0   & 1.00  & 1.00  & 1.00  & 0.98  & 0.00  & 0.00  & 0.00  & 0.00 \\
    &      & 10.0  & 0.98  & 0.77  & 0.97  & 0.43  & 0.01  & 0.01  & 0.00  & 0.01 \\
    &      & 25.0  & 0.34  & 0.01  & 0.46  & 0.00  & 0.00  & 0.00  & 0.02  & 0.01 \\
    &      & 50.0  & 0.01  & 0.01  & 0.05  & 0.00  & 0.04  & 0.01  & 0.01  & 0.00 \\
1.25 & 200 & 5.0   & 0.46  & 0.61  & 0.62  & 0.50  & 0.00  & 0.00  & 0.01  & 0.03 \\
    &      & 10.0  & 0.06  & 0.12  & 0.19  & 0.10  & 0.01  & 0.00  & 0.02  & 0.01 \\
    &      & 25.0  & 0.00  & 0.02  & 0.01  & 0.02  & 0.03  & 0.00  & 0.00  & 0.00 \\
    &      & 50.0  & 0.00  & 0.00  & 0.00  & 0.00  & 0.00  & 0.00  & 0.00  & 0.00 \\
1.25 & 300 & 5.0   & 0.02  & 0.06  & 0.06  & 0.01  & 0.02  & 0.00  & 0.00  & 0.01 \\
    &      & 10.0  & 0.00  & 0.01  & 0.00  & 0.00  & 0.01  & 0.01  & 0.00  & 0.00 \\
    &      & 25.0  & 0.00  & 0.00  & 0.00  & 0.00  & 0.01  & 0.01  & 0.01  & 0.00 \\
& & & & & & & & & &   \\
1.5 & 100  & 5.0   & 1.00  & 1.00  & 1.00  & 1.00  & 0.00  & 0.00  & 0.00  & 0.00 \\
    &      & 10.0  & 1.00  & 1.00  & 1.00  & 1.00  & 0.00  & 0.00  & 0.00  & 0.00 \\
    &      & 25.0  & 0.97  & 0.88  & 0.99  & 0.66  & 0.00  & 0.00  & 0.01  & 0.03 \\
    &      & 50.0  & 0.16  & 0.06  & 0.54  & 0.03  & 0.04  & 0.00  & 0.02  & 0.01 \\
1.5 & 200  & 5.0   & 0.99  & 1.00  & 1.00  & 0.99  & 0.01  & 0.00  & 0.00  & 0.01 \\
    &      & 10.0  & 0.62  & 0.85  & 0.85  & 0.84  & 0.02  & 0.00  & 0.01  & 0.01 \\
    &      & 25.0  & 0.00  & 0.10  & 0.15  & 0.08  & 0.01  & 0.02  & 0.00  & 0.01 \\
    &      & 50.0  & 0.00  & 0.01  & 0.01  & 0.00  & 0.00  & 0.01  & 0.03  & 0.01 \\
1.5 & 300  & 5.0   & 0.49  & 0.61  & 0.59  & 0.53  & 0.01  & 0.01  & 0.00  & 0.00 \\
    &      & 10.0  & 0.02  & 0.17  & 0.17  & 0.09  & 0.00  & 0.00  & 0.00  & 0.02 \\
    &      & 25.0  & 0.00  & 0.00  & 0.00  & 0.00  & 0.00  & 0.00  & 0.01  & 0.00 \\
& & & & & & & & & &   \\
1.75 & 100 & 5.0   & 1.00  & 1.00  & 1.00  & 1.00  & 0.00  & 0.00  & 0.00  & 0.00 \\
    &      & 10.0  & 1.00  & 1.00  & 1.00  & 1.00  & 0.00  & 0.00  & 0.00  & 0.00 \\
    &      & 25.0  & 1.00  & 1.00  & 1.00  & 1.00  & 0.00  & 0.00  & 0.00  & 0.01 \\
    &      & 50.0  & 0.88  & 0.77  & 0.99  & 0.57  & 0.01  & 0.00  & 0.01  & 0.00 \\
1.75 & 200 & 5.0   & 1.00  & 1.00  & 1.00  & 1.00  & 0.00  & 0.00  & 0.00  & 0.00 \\
    &      & 10.0  & 0.97  & 0.99  & 0.99  & 0.99  & 0.01  & 0.01  & 0.00  & 0.01 \\
    &      & 25.0  & 0.27  & 0.60  & 0.72  & 0.59  & 0.01  & 0.00  & 0.00  & 0.01 \\
    &      & 50.0  & 0.00  & 0.10  & 0.14  & 0.07  & 0.00  & 0.02  & 0.02  & 0.01 \\
1.75 & 300 & 5.0   & 0.95  & 0.99  & 1.00  & 0.99  & 0.02  & 0.00  & 0.01  & 0.02 \\
    &      & 10.0  & 0.36  & 0.58  & 0.61  & 0.48  & 0.00  & 0.03  & 0.03  & 0.01 \\
    &      & 25.0  & 0.00  & 0.06  & 0.09  & 0.03  & 0.01  & 0.00  & 0.00  & 0.02 \\
    &      & 50.0  & 0.00  & 0.00  & 0.00  & 0.00  & 0.01  & 0.00  & 0.01  & 0.04 \\
& & & & & & & & & &   \\
2.0 & 100  & 5.0   & 1.00  & 1.00  & 1.00  & 1.00  & 0.00  & 0.00  & 0.00  & 0.00 \\
    &      & 10.0  & 1.00  & 1.00  & 1.00  & 1.00  & 0.00  & 0.00  & 0.00  & 0.00 \\
    &      & 25.0  & 1.00  & 1.00  & 1.00  & 1.00  & 0.00  & 0.00  & 0.00  & 0.00 \\
    &      & 50.0  & 1.00  & 1.00  & 1.00  & 0.99  & 0.00  & 0.00  & 0.01  & 0.01 \\
2.0 & 200  & 5.0   & 1.00  & 1.00  & 1.00  & 1.00  & 0.00  & 0.00  & 0.00  & 0.00 \\
    &      & 10.0  & 1.00  & 1.00  & 1.00  & 1.00  & 0.00  & 0.00  & 0.00  & 0.00 \\
    &      & 25.0  & 0.92  & 0.99  & 0.99  & 0.99  & 0.00  & 0.00  & 0.00  & 0.00 \\
    &      & 50.0  & 0.09  & 0.57  & 0.73  & 0.58  & 0.02  & 0.02  & 0.00  & 0.01 \\
2.0 & 300  & 5.0   & 1.00  & 1.00  & 1.00  & 1.00  & 0.00  & 0.00  & 0.00  & 0.00 \\
    &      & 10.0  & 0.97  & 1.00  & 1.00  & 0.99  & 0.01  & 0.00  & 0.00  & 0.00 \\
    &      & 25.0  & 0.08  & 0.42  & 0.47  & 0.36  & 0.02  & 0.00  & 0.01  & 0.01 \\
    &      & 50.0  & 0.00  & 0.01  & 0.02  & 0.02  & 0.00  & 0.02  & 0.01  & 0.01 \\
2.0 &1000  & 5.0   & 0.00  & 0.01  & 0.01  & 0.00  & 0.01  & 0.01  & 0.00  & 0.02 \\
    &      & 10.0  & 0.00  & 0.00  & 0.00  & 0.00  & 0.01  & 0.01  & 0.00  & 0.01 \\  
 ~   & & & & & & & & & & \\
\hline
\end{tabular}
\end{center}


\label{det_freq}

\end{table*}


\subsection{$\alpha$ versus SDE statistics}
\label{alpha_vs_SDE}
Using the SDE statistic instead of the $\alpha$ statistic, the results 
in terms of detection rates and false alarm percentages obtained by the 
200-harmonic fitting, the 3-spot model and the INL filter are very 
similar in the cases when $\sigma\geq200$ ppm, that is for most of our 
simulations. Specifically, the detection rates differ from those reported 
in Table~\ref{det_freq} by no more than 2-3\%. Only the SBC filter, in a few 
cases, shows an improvement of the detection rate of about 5-6\% but its 
performance is never better than that of the INL filter.

Different results are obtained in the cases where $\sigma=100$ ppm when 
the SDE statistic is adopted: the detection rates of the 3-spot model 
and the SBC filter show a significant improvement reaching those of 
the 200-harmonic fitting with false alarm rates below 2\%. The reason 
of such a behaviour is as follows. The $\alpha$ statistic is much more 
sensitive to single transit events than the SDE statistic, especially 
for low levels of noise. This is due to the fact that, if there is a 
single transit event, the BLS algorithm detects it several times while 
trying different orbital periods, which gives rise to a great dispersion 
of the Signal Residue in the BLS spectrum (see \citealt{Kovacsetal02}, Eq.~5) 
and therefore to a decrease of the SDE statistic that is inversely proportional 
to the standard deviation of the Signal Residue (see \citealt{Kovacsetal02}, Eq.~6). 
The $\alpha$ statistic, on the other hand, does not suffer from this problem and 
therefore it is more suitable to search for monotransits. We prefer to use it 
since in our analysis the period search is carried out between 1 and 150 days. 
However, for the same reason as above, the $\alpha$ statistic is also more 
sensitive to sporadic transit-like features produced by the filtering of 
the stellar variability with the SBC filter and the 3-spot model (see Sect.~\ref{sigma_100} 
and \citealt{BonomoLanza08}, Sect.~5). When we apply the $\alpha$ statistic 
to cases with $\sigma=100$~ppm, those features force us to increase 
the detection thresholds in order to have a false alarm rate below
 the 1\% level in the transitless light curves. Such an increase leads to 
 a worse detection performance in the light curves with transits. From 
 this point of view, the SDE statistic seems to be more robust against 
 false alarms for low noise levels ($\sigma=100$ ppm). Nevertheless, 
 the detection of single transit events is sometimes important. For 
 instance, the CoRoT alarm mode searches for transits while data are 
 arriving and, when it detects a single transit, it allows to change 
 the time sampling of the light curve to seek transit timing variations 
 \citep{Bordeetal03}. In this case, the SDE statistic is not appropriate and 
 the $\alpha$ statistic is certainly more suitable.

\section{Discussion}
\label{disc}
From the above mentioned results we conclude that team 3 of the CoRoT 
Blind Test achieved the best performance mostly thanks to the use of 
the BLS, one of the most powerful transit detection algorithms, rather than 
by virtue of their detrending method. However, the 200-harmonic fitting 
has the advantage of being fully automatic, in particular it is independent 
of the activity level of the star thanks to the high number of free parameters ($N$=401). 
On the other hand, the other filtering techniques need to adjust some of 
their parameters according to the activity level of the star. In particular, 
to fit the light curve of a star more active than the Sun, the 3-spot method 
requires to change the rotation period, the limb-darkening, the maximum spot 
area or the contrast of spots, and the duration of the individually fitted 
time intervals (cf. \citealt{Lanzaetal03} for details). In the case of the INL 
and the SBC filters, the extension of the window should be reduced for stars 
more active than the Sun, otherwise some oscillations or transit-like features 
will appear in the residuals owing to a bad filtering of the variability.

\begin{figure}[!t]
\vspace*{-1.0 cm}
\centering
 \hspace{-0.7 cm}
 \includegraphics[width=9cm]{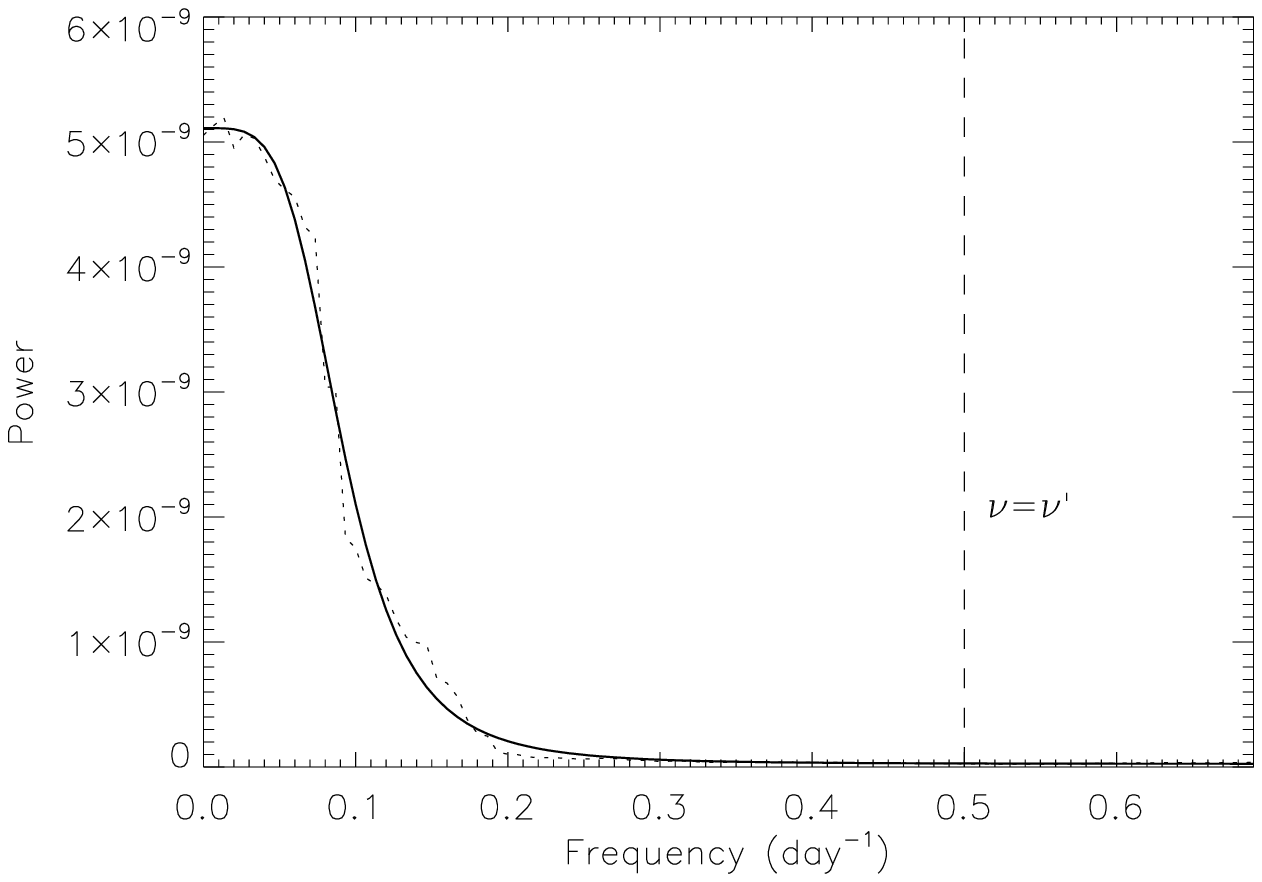} 
 \vspace*{-0.3 cm}
 \caption{\emph{Dotted line}: The power spectrum of the relative variation of one of our simulated light curves; \emph{Solid line}: The best fit performed with a single power law (see Eq.~\ref{powlaw}) with $A$=$5.086\cdot10^{-9}$, $B$=10.92 d, $C$=4.22 and $\mathrm{const}$=$2.449\cdot10^{-11}$. The vertical dashed line indicates the frequency $\nu^{\prime}$, determined by means of Eq.~\ref{freq}, that gives the appropriate extension for the INL and SBC filter windows.}
  \label{fig2}
\end{figure}

We could make the INL and SBC filters automatic by introducing a method 
to choose the extension of the filter window according to the magnetic 
activity level of the star. 
We propose to use a method similar to that suggested by \citet{Reguloetal07}, 
based on the computation of the power spectrum of the data to find the 
frequency intervals where most of the power is concentrated and choose 
automatically the filter window corresponding to the frequency where the 
power spectral density goes below a fixed threshold. 
The power spectrum of the Total Solar Irradiance can be modelled as a sum 
of power laws, each of which corresponds to a separate class of physical 
phenomena, taking place on different characteristic time scales \citep{Harvey85, Aigrainetal04}. 
For our purpose, it is sufficient to perform a best-fit of the 
power spectrum of a light curve by means of a single power law, 
corresponding to the evolution and rotational modulation of active regions, i.e.,
\begin{equation}
f(\nu)=\frac{A}{1+(B \cdot \nu)^{C}} + \mathrm{const},
\label{powlaw}
\end{equation}
where $\nu$ is the frequency, $A$, $B$ and $C$ represent the amplitude, the characteristic time scale and the slope of the power law, respectively (see Fig.~\ref{fig2}). The inverse of the frequency $\nu^{\prime}$ such that 
\begin{equation}
f(\nu^{\prime})-\mathrm{const}=10^{-3} \cdot \max[f(\nu)-\mathrm{const}]
\label{freq}
\end{equation}
gives the window extension to use.

\begin{figure}[!b]
\vspace*{-0.2 cm}
\centering
 \hspace{-0.7 cm}
 \includegraphics[width=9cm]{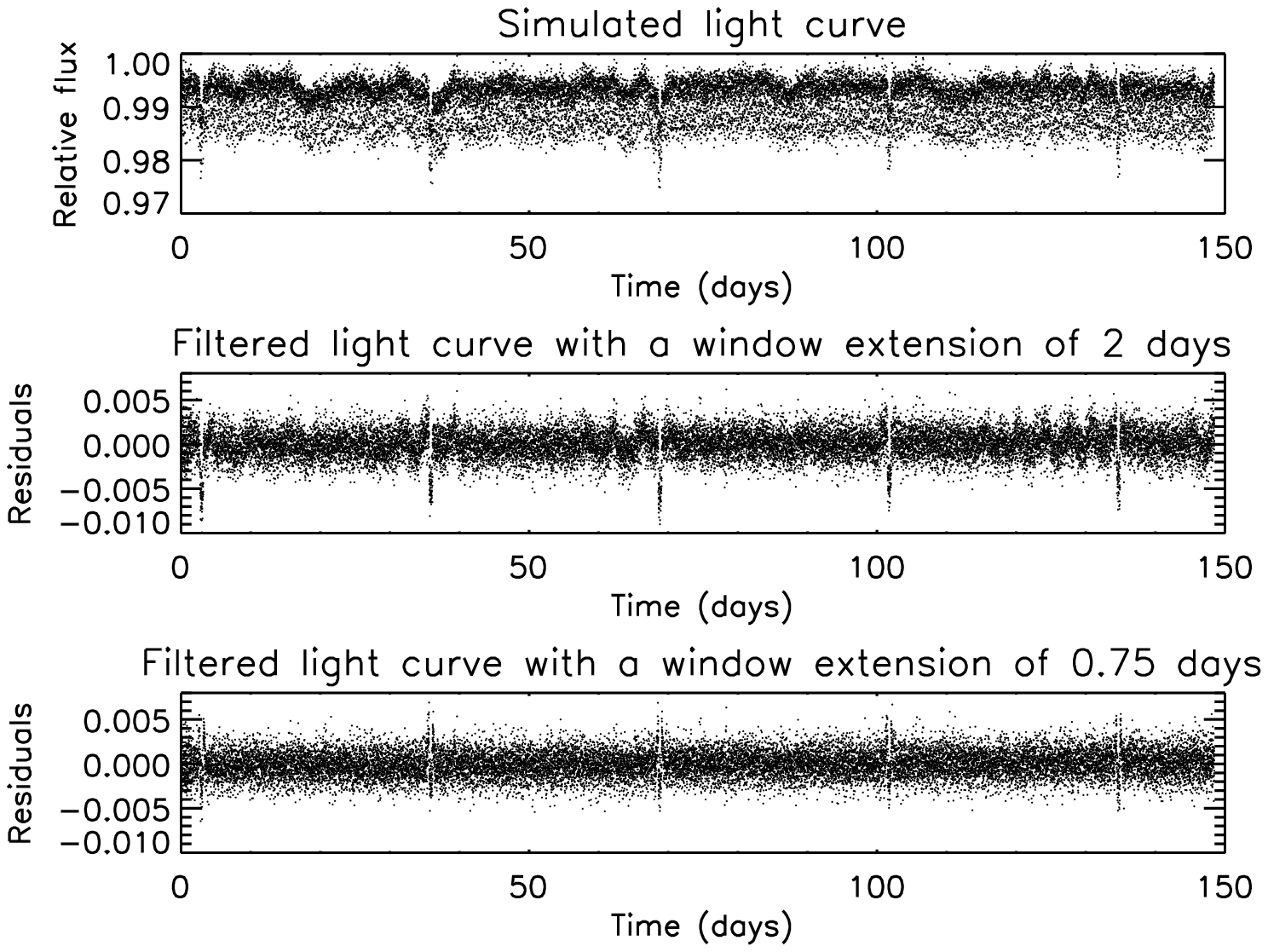} 
 \vspace*{-0.3 cm}
 \caption{\footnotesize
\emph{Upper panel}: one of the light curves with transits simulated for the First CoRoT Blind test (\citealt{Moutouetal05}; ID=460). \emph{Middle panel}: the light curve filtered by means of the INL filter with a window of 2 days. \emph{Bottom panel}: the filtered light curve with a 0.75 day window. Note the disappearance of the transits when the window extension is reduced.}
  \label{fig3}
\end{figure}

An alternative to such a method would be using a very short window extension, i.e., 
only two or three times the transit duration, to filter the light curves of both 
quiet and active stars, as team 5 did for the CoRoT BT1. However, a very short 
window, when not necessary, as in the case of quiet solar-like stars, is not 
advisable because it gives rise to a reduction of the transit depth in the 
filtered light curve (see Fig.~\ref{fig3}) and therefore to a worse detection 
performance when the BLS algorithm is applied.

To show the deterioration of the detection performance with the reduction of the window
extension for the INL and the SBC filters, we consider the simulated light curves for a
planet of $R_{p}$=1.75 R$_{\oplus}$ and analyse them using four different window extensions
of 12, 24, 36 and 48 hours in the same way as explained in Sect.~\ref{sim_anal}. For each
window extension, we redetermine appropriate transit detection thresholds by analysing the
transitless light curves. The results are presented in Table~\ref{det_freq_inl_sbc} where, in
the first column, we list the window extension $\Delta T_{w}$; in the second, the standard
deviation $\sigma$ of the photon noise; in the third, the simulated orbital period $P$ of the
planet; in the fourth and fifth column, the frequency of transit detections $D_{INL}$ and 
$D_{SBC}$ obtained respectively with the INL and the SBC filters; in the sixth and seventh 
column, the false alarm rates $FA_{INL}$ and $FA_{SBC}$ produced by both methods.


\begin{table*}
\caption{Fraction of positive detections and false alarms obtained by the INL and the SBC filters for different extensions of the filter window $\Delta T_{w}$ and simulated light curves with transits of a 1.75 R$_{\oplus}$ planet.}
\vspace*{-2mm}
\begin{minipage}[t]{5.0cm}
\begin{center}
\begin{tabular}{|ccccccc|}
 \hline
& & & & & & \\
$\Delta$ $T_{\rm w}\mbox{~~}$ & $\sigma\mbox{~~~}$ & $P$ & $D_{\rm INL}$ & $D_{\rm SBC}$ & $FA_{\rm INL}$ & $FA_{\rm SBC}$ \\
  (h) & (ppm) & (d) & & & &\\ 
& & & & & & \\
\hline
& & & & & & \\
& & & & & & \\
12  & 100  & 5.0   & 1.00  & 1.00  & 0.00  & 0.00   \\
    &      & 10.0  & 1.00  & 1.00  & 0.00  & 0.00   \\
    &      & 25.0  & 1.00  & 1.00  & 0.01  & 0.00   \\
    &      & 50.0  & 0.09  & 0.21  & 0.03  & 0.00   \\
12  & 200  & 5.0   & 1.00  & 1.00  & 0.00  & 0.00   \\
    &      & 10.0  & 0.89  & 0.94  & 0.01  & 0.01   \\
    &      & 25.0  & 0.06  & 0.08  & 0.02  & 0.00   \\
    &      & 50.0  & 0.00  & 0.00  & 0.00  & 0.01   \\
12  & 300  & 5.0   & 0.86  & 0.92  & 0.03  & 0.03   \\
    &      & 10.0  & 0.19  & 0.26  & 0.03  & 0.03   \\
    &      & 25.0  & 0.00  & 0.00  & 0.02  & 0.02   \\
    &      & 50.0  & 0.00  & 0.00  & 0.02  & 0.00   \\
12  & 1000 & 5.0   & 0.00  & 0.00  & 0.01  & 0.01   \\ 
& & & & & & \\
& & & & & & \\
24  & 100  & 5.0   & 1.00  & 1.00  & 0.00  & 0.00   \\
    &      & 10.0  & 1.00  & 1.00  & 0.00  & 0.00   \\
    &      & 25.0  & 1.00  & 0.99  & 0.00  & 0.01   \\
    &      & 50.0  & 0.98  & 0.98  & 0.02  & 0.00   \\
24  & 200  & 5.0   & 1.00  & 1.00  & 0.00  & 0.00   \\
    &      & 10.0  & 0.98  & 0.99  & 0.01  & 0.01   \\
    &      & 25.0  & 0.47  & 0.41  & 0.01  & 0.00   \\
    &      & 50.0  & 0.06  & 0.01  & 0.01  & 0.00   \\
24  & 300  & 5.0   & 0.96  & 0.97  & 0.01  & 0.02   \\
    &      & 10.0  & 0.44  & 0.50  & 0.02  & 0.00   \\
    &      & 25.0  & 0.03  & 0.03  & 0.01  & 0.02   \\
    &      & 50.0  & 0.00  & 0.00  & 0.00  & 0.02   \\
24  & 1000 & 5.0   & 0.00  & 0.00  & 0.01  & 0.01   \\
& & & & & & \\
\hline
\end{tabular}
\end{center}
\end{minipage}
\hspace*{2.8cm}
\begin{minipage}[t]{5.0cm}
\begin{center}
\begin{tabular}{ccccccc|}
\hline
& & & & & & \\
$\Delta$ $T_{\rm w}\mbox{~~}$ & $\sigma\mbox{~~~}$ & $P$ & $D_{\rm INL}$ & $D_{\rm SBC}$ & $FA_{\rm INL}$ & $FA_{\rm INL}$ \\
  (h) & (ppm) & (d) & & & & \\ 
& & & & & & \\
\hline
& & & & & & \\
& & & & & & \\

36  & 100  & 5.0   & 1.00  & 1.00  & 0.00  & 0.00   \\
    &      & 10.0  & 1.00  & 1.00  & 0.00  & 0.00   \\
    &      & 25.0  & 1.00  & 0.98  & 0.00  & 0.02  \\
    &      & 50.0  & 0.99  & 0.80  & 0.01  & 0.01   \\
36  & 200  & 5.0   & 1.00  & 1.00  & 0.00  & 0.00   \\
    &      & 10.0  & 0.99  & 0.99  & 0.00  & 0.01   \\
    &      & 25.0  & 0.71  & 0.46  & 0.01  & 0.00   \\
    &      & 50.0  & 0.12  & 0.03  & 0.02  & 0.00   \\
36  & 300  & 5.0   & 0.99  & 0.96  & 0.01  & 0.02   \\
    &      & 10.0  & 0.57  & 0.43  & 0.02  & 0.00   \\
    &      & 25.0  & 0.06  & 0.02  & 0.00  & 0.01   \\
    &      & 50.0  & 0.00  & 0.00  & 0.00  & 0.02   \\
36  & 1000 & 5.0   & 0.00  & 0.00  & 0.00  & 0.01   \\
& & & & & & \\
& & & & & & \\
48  & 100  & 5.0   & 1.00  & 1.00  & 0.00  & 0.00   \\
    &      & 10.0  & 1.00  & 1.00  & 0.00  & 0.00   \\
    &      & 25.0  & 1.00  & 1.00  & 0.00  & 0.01   \\
    &      & 50.0  & 0.99  & 0.57  & 0.01  & 0.00   \\
48  & 200  & 5.0   & 1.00  & 1.00  & 0.00  & 0.00   \\
    &      & 10.0  & 0.99  & 0.99  & 0.00  & 0.01   \\
    &      & 25.0  & 0.72  & 0.59  & 0.00  & 0.01   \\
    &      & 50.0  & 0.14  & 0.07  & 0.02  & 0.01   \\
48  & 300  & 5.0   & 1.00  & 0.99  & 0.01  & 0.02   \\
    &      & 10.0  & 0.61  & 0.48  & 0.03  & 0.01   \\
    &      & 25.0  & 0.09  & 0.03  & 0.00  & 0.02   \\
    &      & 50.0  & 0.00  & 0.00  & 0.01  & 0.04   \\
48  & 1000 & 5.0   & 0.00  & 0.00  & 0.00  & 0.01   \\
& & & & & & \\
\hline 
\label{det_freq_inl_sbc}
\end{tabular}
\end{center}
\end{minipage}
\end{table*}


Let us consider for example the case with a window of 12~h, $\sigma$=300 ppm and $P$=10 d. 
In this case the window extension is three times the transit duration. 
We note that: a) after the INL filtering, the positive detections are 19\% 
versus 61\% when a 2 day (48 h) window was applied; b) the positive detections 
obtained after SBC filtering are 26\% versus 48\% with a window of 2 days. Therefore, 
if we used a window extension equal to three times the transit duration 
for the INL and the SBC filters, the 3-spot model would be the best method 
(with its 58\% positive detections, see Table~\ref{det_freq}) and even the 
performance of the 200-harmonic fitting (36\% positive detections) would 
be better than that of the INL and SBC filters. 

Similar results are obtained in the case with a window extension 
of 24~h, $\sigma$=200 ppm and $P$=25 d. In this case the window is four 
and a half times the transit duration. In the case of the INL filtering, 
the positive detections decrease from 72\% with a window of 2 days to 47\% 
with a 1 day window; regarding the SBC filter, they decrease from 59\% to 41\%. 
Once again, the 3-spot model would prove to be the best filtering method (see Table~\ref{det_freq}).

Note that in some cases (i.e., $\sigma$=300 ppm, $P$=10 d) the performance 
of the SBC filter does not vary in a monotonic way as a function of the 
window extension, owing to the effect of the changing detection threshold.

Looking at the results obtained with a 12~h window, we see that the 
SBC filter is slightly more efficient than the INL filter. This is connected 
with the fact that, reducing the filter window, we reduce the number 
of data points from which each value of the continuum is computed. With a fewer 
data points, the median used by the INL filter is more affected by statistical 
fluctuations than the arithmetic mean used by the SBC, so the former gives 
rise to a distribution of the residuals with a greater standard deviation and 
higher tails than that obtained with the SBC filter. In other words, in the 
case of a short window with a small number of data points, the continuum is 
better computed by the SBC filter, giving rise to a better detection performance 
by means of the BLS.

Although we have pointed out that, with an \-i\-nap\-pro\-pria\-te window, 
the INL filter has a performance worse than the \,\,\,\,\,\,\,\,\,\,\,\,\,\,3-spot model, 
the latter is computationally much more intensive, being based on a physical 
model of stellar variability. The time it takes to filter one light curve is $\sim$10 min 
against $\sim$3 min with the 200-harmonic fitting and just a few seconds with the INL 
and the SBC filters. Therefore, the use of the INL filter with a window extension 
determined according to the magnetic activity level of the star is preferable to 
the 3-spot model.

\section{Conclusions}
We have performed extensive numerical experiments to compare the performance 
of four different variability filters for the detection of Earth-like planetary 
transits by means of a box-shaped transit finder algorithm. The INL filter 
has proved to be the best method to filter light curves of quiet solar-like 
stars when a sufficiently long window can be chosen. We have shown that the 
choice of the window of the filter is  critical since its performance 
depends significantly on it. We point out that the window must be as long 
as possible, according to the magnetic activity level of the star. A method 
to choose the extension of the window, similar to that proposed by 
\citet{Reguloetal07}, is shown in Sect.~\ref{disc}. The INL filter, when used 
with a sui\-ta\-ble choice of its window, has a better performance than 
more complicated and \-com\-pu\-ta\-tio\-nal\-ly intensive methods of fitting 
solar-like variability, like the 200-harmonic fitting or the 3-spot model.

\begin{acknowledgements}
The authors are grateful to Dr.~R.~Alonso for useful discussions and acknowledge an anonymous Referee for valuable comments. They are grateful also to Drs.~U.~Becciani, A.~Costa,  A.~Grillo, 
and  the system managers of the Trigrid and Cometa Consortia for their technical advice and 
kind assistance \-du\-ring the implementation and running of the numerical experiments on grid-based 
high performance computing systems.  
The \-a\-vai\-la\-bi\-li\-ty of unpublished data of the VIRGO Experiment
on the coo\-pe\-ra\-ti\-ve ESA/NASA Mission SoHO from the VIRGO Team through PMOD/WRC, Davos,
Switzerland, is gratefully acknowledged. 
ASB and AFL gratefully acknowledge support from the Italian Space Agency (ASI) under contract  ASI/INAF I/015/07/0, work package 3170. Part of this work was carried out during a visit to Exeter by ASB, for which the authors gratefully acknowledge support from grant PP/D001617/1 from the UK Science and Technology Facilities Council.
This research has made use of results produced by the PI2S2 Project managed by the Consorzio COMETA, 
a project co-funded by the Italian {\it Ministero dell'Istruzione, Universit\`a e Ricerca} (MIUR) within the 
{\it Piano Operativo Nazionale "Ricerca Scientifica, Sviluppo Tecnologico, Alta Formazione" (PON 2000-2006)}. 
More information is available at http://www.pi2s2.it and http://www.consorzio-cometa.it.

Active star research and exoplanetary studies at INAF-Catania Astrophysical Observatory and the Department of Physics and Astronomy of Catania University are funded by MIUR ({\it Ministero dell'Istruzione, Universit\`a e Ricerca}), and by {\it Regione Siciliana}, whose financial support is gratefully
acknowledged. 
This research has made use of the ADS-CDS databases, operated at the CDS, Strasbourg, France.
\end{acknowledgements}



\begin{thebibliography}{}
\bibliographystyle{aa}

\bibitem[\protect\citeauthoryear{Aigrain et al.}{2004}]{Aigrainetal04}
Aigrain, S., Favata, F., Gilmore, G. 2004, \aap, 414, 1139

\bibitem[\protect\citeauthoryear{Aigrain \& Irwin}{2004}]{AigrainIrwin04}
Aigrain, S., Irwin, M. 2004, \mnras, 350, 331 

\bibitem[\protect\citeauthoryear{Baglin}{2003}]{Baglin03} 
Baglin, A. 2003, Adv. Sp. Res., 31, 345

\bibitem[\protect\citeauthoryear{Becciani}{2007}]{Becciani07}
Becciani, U. 2007, in Grid-enabled Astrophysics, L.~Benacchio, F. Pasian, Eds., Edizioni  Polimetrica; 
p. 179

\bibitem[\protect\citeauthoryear{Bennett et al.}{2008}] {Bennettetal08}
Bennett, D. P., Bond, I. A., Udalski, A., Sumi, T., Abe, F. et al. 2008, \apj, in press, (arXiv:0806.0025)

\bibitem[\protect\citeauthoryear{Bonomo \& Lanza} {2008}]{BonomoLanza08}
Bonomo, A. S., Lanza, A. F. 2008, \aap, 482, 341

\bibitem[\protect\citeauthoryear{Bord\'e et al.}{2003}]{Bordeetal03}
Bord\'e, P., Rouan, D., L\'eger, A. 2003, \aap, 405, 1137 

\bibitem[\protect\citeauthoryear{Borucki et al.}{2004}]{Boruckietal04} 
Borucki, W., Koch, D., Boss, A., et al. 2004, 
in Second Eddington Workshop: Stellar structure and habitable planet finding, eds. F. Favata, S. Aigrain, \& A. Wilson, ESA SP-538, 177 

\bibitem[\protect\citeauthoryear{Carpano et al.}{2003}]{Carpanoetal03}
Carpano, S., Aigrain, S., Favata, F. 2003, \aap, 401, 743

\bibitem[\protect\citeauthoryear{Carpano \& Fridlund}{2008}]{CarpanoFridlund08}
Carpano, S. \& Fridlund, M. 2008, \aap, 485, 607

\bibitem[\protect\citeauthoryear{Defa\"y et al.}{2001}]{Defayetal01}
Defa\"y C., Deleuil, M., Barge, P. 2001, \aap, 365, 330 

\bibitem[\protect\citeauthoryear{Doyle et al.}{2000}]{Doyleetal00}
Doyle, L. R., Deeg, H. J., Kozhevnikov, V. P., Oetiker, B., Mart\'{\i}n, E. L., et al.
2000, \apj, 535, 338 

\bibitem[\protect\citeauthoryear{Fr\"ohlich \& Lean} {2004}]{FrohlichLean04}
Fr\"ohlich, C., Lean, J. 2004, A\&ARv, 12, 273

\bibitem[\protect\citeauthoryear{Harvey}{1985}]{Harvey85}
Harvey, J. W. 1985, in ESA Future missions in solar, heliospheric and space plasma physics,
eds. E. Rolfe, \& B. Battrick, ESA SP-235, 199

\bibitem[\protect\citeauthoryear{Jenkins}{2002}]{Jenkins02}
Jenkins, J. M. 2002, \apj, 575, 493 

\bibitem[\protect\citeauthoryear{Jenkins et al.}{2002}]{Jenkinsetal02}
Jenkins, J. M., Caldwell, D. A., Borucki, W. J. 2002, \apj, 564, 495  

\bibitem[\protect\citeauthoryear{Kov\'acs et al.}{2002}]{Kovacsetal02}
Kov\'acs, G., Zucker, S., Mazeh, T. 2002, \aap, 391, 369 

\bibitem[\protect\citeauthoryear{Lanza et al.}{2003}]{Lanzaetal03}
Lanza, A. F., Rodon\`o, M., Pagano, I., Barge, P., Llebaria, A. 2003, \aap, 403, 1135

\bibitem[\protect\citeauthoryear{Lanza et al.}{2007}]{Lanzaetal07}
Lanza, A. F., Bonomo, A. S., Rodon\`o, M. 2007, \aap, 464, 741 

\bibitem[\protect\citeauthoryear{Mayor et al.}{2008}] {Mayoretal08}
Mayor, M., Udry, S., Lovis, C., Pepe, F., Queloz, D., et al. 2008, \aap, submitted (arXiv:0806.4587) 

\bibitem[\protect\citeauthoryear{Morse \& Feshbach}{1954}]{MorseFeshbach54}
Morse, P. M., Feshbach, H. 1954, Methods of Theoretical Physics, McGraw-Hill Book Co. Inc., New York

\bibitem[\protect\citeauthoryear{Moutou et al.}{2005}]{Moutouetal05} 
Moutou, C., Pont, F., Barge, P., Aigran, S., Auvergne, M., et al. 2005, \aap, 437, 355

\bibitem[\protect\citeauthoryear{Regulo et al.}{2007}] {Reguloetal07}
Regulo, C., Almenara, J. M., Alonso, R., Deeg H., Roca Cortes T. 2007, \aap, 467, 1345

\bibitem[\protect\citeauthoryear{Wolszczan \& Frail}{1992}] {WolszczanFrail92}
Wolszczan, A., Frail, D. A. 1992, Nature, 335, 145

\bibitem[\protect\citeauthoryear{Wolszczan}{1994}] {Wolszczan94}
Wolszczan, A., 1994, Science, 264, 538




\end{thebibliography}
\end{document}